\documentclass{article}
\usepackage{arxiv}
\usepackage[utf8]{inputenc} 
\usepackage[T1]{fontenc}    
\usepackage{hyperref}
\hypersetup{colorlinks,allcolors=blue}
\usepackage{url}            
\usepackage{booktabs}       
\usepackage{amsfonts}       
\usepackage{nicefrac}       
\usepackage{microtype}      
\usepackage{lipsum}
\usepackage{graphicx}
\graphicspath{ {./images/} }
\usepackage{sectsty}
\usepackage[compact]{titlesec}
\usepackage{hyperref}
\hypersetup{
    colorlinks,
    linkcolor=blue,
    filecolor=blue,      
    urlcolor=blue,
}
\usepackage{dirtytalk}
\usepackage{longtable}
\usepackage[toc,page]{appendix}
\usepackage[tableposition=top]{caption}
\usepackage[symbol]{footmisc}
\title{Abstracts}

\title{Measurement in AI Policy: \\
Opportunities and Challenges\footnote[1]{\scriptsize We are grateful to the financial support of Stanford Institute for Human-Centered Artificial Intelligence (HAI) for hosting the roundtable workshop; the AI Index sponsors McKinsey \& Company, Google, OpenAI, Genpact, AI21 labs, PricewaterhouseCoopers (PwC); the organizing committee including Susan Athey, Erik Brynjolfsson, Juan Carlos-Niebles, John Etchemendy, Barbara Grosz, Fei-Fei Li, Terah Lyons, James Manyika, Michael Sellitto, Yoav Shoham; special thank you to the breakout session moderators including Eileen Donahoe, Karine Perset, Margarita Quihuis; We are grateful to Deep Ganguli and Daniel Zhang for invaluable feedback on the draft; Special thank you to the session panelists including Monica Anderson, Alessandro Annoni, Mohsen Bayati, Guy Berger, Tulsee Doshi, Timnit Gebru, Ramesh Johari, Henry Kautz, Lars Kotthoff, Jamila Loud, Ashley Llorens, Christos Makridis, Dewey Murdick, Christopher Potts, Anand Rao, Dorsa Sadigh, Jake Silberg, Andrew Smart, Prasanna (Sonny) Tambe, Rachel Thomas, Christopher Walker, Susan Woodward, Chenggang Xu, James Zou, and all the presenters and participants; Celia Clark, Agata Foryciarz, Lisa Simon, and Tamara Prstic for help organizing the breakout groups; Ruth Starkman, Peter Cihon, and Ankita Banerjea for help on the copy edits.}}

\author{
Saurabh Mishra
 \\
AI Index, HAI \\ Stanford University \\
\texttt{saurabh.mishra@stanford.edu} \\
   \And
Jack Clark \\
 AI Index, HAI \\ OpenAI \\
\texttt{jack@openai.com} \\
 \And
C. Raymond Perrault \\
AI Index, HAI \\ SRI  International \\
\texttt{ray.perrault@sri.com} \\
}

\begin{document}
\maketitle

\begin{abstract}
As artificial intelligence increasingly influences our world, it becomes crucial to assess its technical progress and societal impact. This paper surveys problems and opportunities in the measurement of AI systems and their impact, based on a workshop held at Stanford University in the fall of 2019. We identify six summary challenges inherent to measuring the progress and impact of AI, and summarize  over 40 presentations and associated discussions from the workshop. We hope this can inspire research agendas in this crucial area. 
\end{abstract}

\keywords
\ {artificial intelligence,  \ policy, \ performance evaluation, \ risk management, \ standards, \ economic inequality, \ sustainable development, \ geopolitics, \ fairness, \ ethics}

\section{Introduction}
The rapid development of Artificial Intelligence and its growing effects on science, the economy and society have increased the need to better assess its technical progress, and economic and societal impact.  

On October 30, 2019, the \href{aiindex.org}{AI Index} \cite{shoham2017towards, shoham2018ai, perrault2019ai} held a workshop at the Stanford Institute for \href{https://hai.stanford.edu}{Human-Centered Artificial Intelligence (HAI)} convening over 150 interdisciplinary experts to discuss issues in the measurement and assessment of artificial intelligence.\footnote[1]{\footnotesize We are grateful to the financial support of Stanford Institute for Human-Centered Artificial Intelligence (HAI) for hosting the roundtable workshop; the AI Index sponsors McKinsey \& Company, Google, OpenAI, Genpact, AI21 labs, PricewaterhouseCoopers (PwC); the organizing committee including Susan Athey, Erik Brynjolfsson, Juan Carlos-Niebles, John Etchemendy, Barbara Grosz, Fei-Fei Li, Terah Lyons, James Manyika, Michael Sellitto, Yoav Shoham; We are grateful to Deep Ganguli and Daniel Zhang for invaluable feedback on the draft; Special thank you to the breakout session moderators including Eileen Donahoe, Karine Perset, Margarita Quihuis; the session panelists including Monica Anderson, Alessandro Annoni, Mohsen Bayati, Guy Berger, Tulsee Doshi, Timnit Gebru, Ramesh Johari, Henry Kautz, Lars Kotthoff, Jamila Loud, Ashley Llorens, Christos Makridis, Dewey Murdick, Christopher Potts, Anand Rao, Dorsa Sadigh, Jake Silberg, Andrew Smart, Prasanna (Sonny) Tambe, Rachel Thomas, Christopher Walker, Susan Woodward, Chenggang Xu, James Zou, and all the presenters and participants; Celia Clark, Agata Foryciarz, Lisa Simon, and Tamara Prstic for help organizing the breakout groups; Ruth Starkman, Peter Cihon, and Ankita Banerjea for help on the copy edits.} Presentations and discussion sessions covered areas like technical progress and measurement, engineering, computational statistics, economics, law, policy, management science, and human rights. Participants represented a variety of institutions,  including academic, private, non-government, and public sector organizations. A full list of participants and their organizations are included in the Appendix. Workshop resources including detailed summary discussions and presentations are available publicly on \href{https://drive.google.com/drive/folders/1vfYhxJVtvlmo5EkxM7b0vwc9dH3n75nd?usp=sharing} {google drive}.  

Across more than 40 workshop presentations, we identified six central problems that researchers face measuring the progress and impact of AI systems. These problems and some of the questions the researchers inspire are:
\newpage

\paragraph{1) What is AI? }
\

How do we define AI? This seemingly simple question exacts significant downstream effects on assessment of R\&D productivity, public and private investment, labor market impact, and many more areas. A broadly accepted definition would facilitate collaboration among organizations collecting AI-related data across sectors and geography.

\paragraph{2) What contributes to AI progress? }
\

Technical progress in AI has mainly been measured --- and driven --- by comparing the performance of different algorithms according to different metrics across publicly-available challenge datasets.  Working on a single dataset may lead to overfitting and overstatement of progress. Bias in datasets can lead to biased predictions and the illusion of progress. Easier challenge datasets get abandoned in favor of more difficult ones, which makes tracking progress over long periods of time more difficult. Properties of algorithms beyond their accuracy are of increasing interest, such as robustness to statistical variation, transferability to new domains, size of training datasets (or generalizability to small training data) and compute requirements.  What should be the measure of progress and how do we explore the tradeoffs between them and accuracy?

\paragraph{3) How do we use and improve bibliometric data to analyze AI and its impact on the world?   }
\

Bibliometric data has long been essential to assess the relative contributions by individual researchers, institutions, and countries to a given scientific field, as well as mapping their collaborations. This data can be used to assess the gender and demographic contributions of those who work in the field of AI. Bibliometrics data can also help us map out the relationship between AI and broader investments in business, science research funding, and education. How might more widely accepted definitions of AI impact bibliographic practices?  Can we use bibliometrics data and analysis to map the lifecycle from invention and the real world application of AI techniques?  Can we develop more accurate methods to automate the analysis of authors' names that contribute to AI research reliably by geography, gender, and other entity attributes?

\paragraph{4) How can we measure the economic impact of AI, especially labor market dynamics but also interaction with economic growth and well-being?}
\

AI has distributional implications on the economy and its impacts need to be more carefully assessed and analyzed. To assess how AI contributes to inequality, it is  important to measure AI inputs, i.e., skills, software, data, and management practices as well as AI outputs i.e. AI consumption patterns. Both are difficult to measure as they are often service-based and their intangibility remains a measurement challenge. This challenge becomes even more difficult in developing countries. In both cases, we confront the same questions: How is the deployment of AI impacting the working lives of manual workers with routine tasks? What about workers in white-collar, high-skill jobs? How can we know which organizations are actively deploying AI technologies into their businesses and institutions? How do we measure the human and technical supply chain of AI systems? Can we conduct broader micro and macroeconomic research and case studies of AI technology to better understand causal patterns and hidden interactions between AI growth and well-being?

\paragraph{5) How can we measure the societal impact of AI, particularly on sustainable economic development and the potential risks of AI to diversity, human rights, and security?}

\ 

AI has the potential to address societal challenges spanning all 17 of the United Nations Sustainable Development Goals (SDGs). Already, 169 metrics have been proposed to track success on human rights and the UN SDGs \cite{chui2018notes}. Large-scale use of AI for social good must address explainability of AI decisions, bias, data privacy and security, and consider the use (and abuse) of AI by various actors. Today, we have limited visibility as to where applications of AI for social good are deployed, in which domains, and by which organizations. There is a paucity of structured data for AI to address the SDGs.

\paragraph{6) How can we measure the risks and threats of deployed AI systems?}
\ 

How can we better assess the threats --- both current and potential --- of AI systems? What tools can we use to analyze the risks of fake news, deep fake videos, autonomous weapons, surveillance, applications that violate privacy, and so on? What metrics could help us track each of these issues? How can we analyze the impacts of AI and use this data to help us intervene where it has severe societal consequences? 

What roles do researchers play in helping to design AI systems for furthering social good? Can AI tools be used to address inequality or to  manage geopolitical risks? How could we develop governance systems that could help AI support  human autonomy? And how can we study both the macro-scale impacts of AI, as well as its specific impact on individuals?

We arrived at these six central problems from in depth discussion in three breakout sessions on the following broad topic areas:

\paragraph{1. Research and Development, including Technical Performance.} This group reviewed data related to journal publications, conferences, and patents,  as well as questions in the measurement of the performance of AI algorithms.

\paragraph{2. Economic Impact and Societal Considerations for Policy Decisions.} This group  reviewed resource allocation questions related to public and private investment including challenges for skill-reallocation, economic loss, including  job loss, qualified labor force reduction, distributional challenges, income inequality, and opportunities for economic diversification.

\paragraph{3. AI for Sustainable Development and Human Rights: Inclusion, Diversity, Human Dignity. } This group reviewed data and measurements indicating the positive potential of AI to serve the SDGs. Alongside these optimistic inquiries, this group also investigated the risks of AI in areas such as privacy, vulnerable populations, human rights, workplace and organizational policy. The socio-political consequences of AI raise many complex questions that require continued rigorous examination.

The remainder of the paper details findings from the conference in a coarse to fine manner. In particular, in Section 2, we provide a top-level summary of the plenary talk (by Susan Athey on the importance of measurement in AI) and a synthesis of all talks from the above three breakout groups. In Section 3, we provide a detailed summary for each individual talk from each breakout session.

\

\section{Summary of Plenary Presentation and Sub-Groups}
\label{sec:headings1}

This section presents the summary of the plenary presentation and the three workshop sub-groups.

\subsection{\textit{Summary of Plenary Presentation}} 

\textbf{\textit{\large{Measurement meets AI}}}\\ 
\textbf{\large{Susan Athey, Stanford University}}

Susan Athey kicked off the day with a presentation about the importance of getting measurement right, talking about key considerations in choosing measurements and what good measurements can look like. She drew parallels between the AI Index project and the guidelines published by the American Economic Association (AEA) on the principles of economic measurement, that underscore the importance of measuring the economy consistently and reliably for all stakeholders including central bankers, researchers, investors, and  governments. In a constantly changing world, providing meaningful and interpretable measurements in a timely manner remains challenging, especially measuring shifts in the economy and AI. Key measurement challenges from the presentation included:

- How to define metrics that are measurable in the short term but are related to long-term outcomes. While what one would like to measure may be distant outcomes such as earnings after an education intervention, researchers cannot wait ten years for those outcomes to become available to evaluate the experiment. Short-term measures that are good predictors of longer-term outcomes are therefore  extremely useful  in social impact research.

- Another frequent measurement challenge is whether organizations are setting the right target measurement as a goal. Sometimes improving short-term measures can be directly juxtaposed to improving long-term outcomes, so it is important to be mindful of what one wants to ultimately maximize.  

Researchers should continue exploring interdisciplinary approaches. To get the measurement in AI policy right,  it is important to understand the domain, the principles of measurement as well as the technology, which requires people with different backgrounds and expertise. Measuring AI well across countries and industries will identify what technologies are working, which in turn attracts more funds to projects that are effective.

\subsection{\textit{Summary of Sub-Group 1: Research and Development (including Technical Performance)}}

This session focused on how to measure and assess technical progress and impact of AI research. Some of the core problems the session focused on included:

- How to define AI and to correctly capture relevant bibliometric data. How to define “AI” as a field. AI research is highly interdisciplinary with competing definitions of the term "AI", which complicates measuring progress in AI research and distinguishing it from growth in other, related areas. Some of the workshop's solutions focused on identifying and analyzing progress in sub-domains (such as computer vision, natural language processing). Maria de Kleijn-Lloyd noted that to truly describe developments of a given domain, it is important not to filter out information closely related to but not captured by our definition.

- Can we use quantitative metrics to automatically identify AI research that proposes new ideas, versus AI research that extends existing methods? The attendees also considered ways to measure breakthroughs and contributions from individual publications to progress on particular AI tasks. Lars Kotthoff presented a method for quantitatively analyzing contributions of publications to progress on tasks, and drew the distinction between research that generates new ideas and research that presents complementary approaches. During the discussion, attendees considered the importance of measuring impact, and agreed that they would like to see work leads to more than published improvements, and that marginal improvements often constitute important contributions.

- What technical tools exist today to help us search over literature corpora (e.g., the AI Index's \href{https://arxiv.aiindex.org}{arXiv Monitor}), and the relative strengths and weaknesses of these tools, particularly to benchmark technical performance at scale. The volume and speed of publishing in the field make it challenging to continuously track progress, even on individual tasks. Kostas Stathoulopoulos, Robert Stojnic, and Matthew Kenney presented work aimed at enabling the analysis of arXiv submissions. They noted that submission formats vary and the variety of data sources, pretraining, preprocessing, training and postprocessing methods makes it difficult to compare models published even as improvements on the same benchmark. There is now a significant effort to streamline the process of making these comparisons.

- There are tradeoffs between focusing on a single metric to measure model performance versus diverse metrics to evaluate the capabilities of a system. Many participants brought up the contrast between the way progress is often reported --- measured by improvement on a single metric, and meaningful progress on a task. Improvement on a single metric may be a result of overfitting to that metric without achieving meaningful progress, and may overstate actual capacity to do things in the real world. Christopher Potts brought up the issue of comparisons to human performance, noting that they often underestimate human performance, while methods that achieve high performance on constrained tasks often fail on mildly adversarial examples. Dorsa Sadigh, Ashley Llorens, and Andrew Lohn noted the importance of extensive testing and using multiple metrics to ensure the safety of automated systems. Mohsen Bayati noted that the mismatch between good performance on a metric and the adaptation to real-world constraints is one of the big reasons why many biomedical informatics methods have not seen wide adaptation in healthcare, despite publications claiming promising results.

- How growth in computational power is leading to measurable improvements in AI capabilities, while also raising issues of the efficiency of AI models and the inherent inequality of compute distribution. Fairness, inclusion, and ethics were also brought up during the discussions. Dorsa Sadigh and Matthew Kenney mentioned that despite growing researchers’ interest in exploring ethical dimensions of their fields, the topic is absent from main conference proceedings, and relegated to small workshops. Finally, the participants discussed unequal access to computational power --- which has been a large contributor to recent improvements in model performance in fields such as Natural Language Processing --- and the environmental cost associated with building highly computationally complex models.

\subsection{\textit{Summary of Sub-Group 2: Economic Impact and Societal Considerations for Policy Decisions}}
\label{sec:others2}

This session focused on the economic and societal related analysis to guide policy discussions. Some of the core problems discussed included:

- A recurring theme of the session was the fundamental question of what is AI. Because of the speed of development of AI and its evolving nature, we struggled with definitional issues at an operational level; from what is AI, to who is an AI practitioner and which skills do developing and operating an AI system require; to what is an AI company. 

- Another discussion area was measuring the adoption of AI. In addition to scientific research, the participants discussed measurement challenges for national or institutional metrics on AI human capital, training supply, data, patents, AI entrepreneurship and industry, software development, as well as AI public policy initiatives and public opinion. Susan Woodward from Sandhill Econometrics asked what are AI companies i.e. what companies produce AI tools, use AI tools, or both. Comparing AI-producing companies in the US and China, Chenggang Xu reported that Chinese AI companies tend to be medium-sized while US AI companies are, with the exception of chipmakers, small. He also found that most AI projects in both China and the US use open AI platforms by US-based companies Google and Facebook. 

- Several speakers highlighted the difficulty of measuring AI inputs as well as AI outputs. AI inputs --- skills, software, data, and management practices --- are rapidly changing and often intangible, not transacted through traditional physical markets so their value is not assessed. Dan Rock from MIT noted that the high fixed costs and low marginal costs of AI mean that investment-side measures may not accurately reflect value. Prasanna (Sonny) Tambe emphasized that AI outputs are also difficult to measure as they are often service-based, with a large gap between investment and use.  

- A number of researchers are using AI labor demand i.e. firms’ jobs data as a proxy for AI adoption to complement slower and more expensive survey data. Others are using AI mentions in earning calls and AI skills data. Bledi Taska reported that according to jobs data, AI is rapidly transforming industries such as information/high tech, services, finance and insurance, manufacturing (but much less slowly), healthcare, and transportation. Evan Schnidman reported that AI mentions in earning calls by sector are highest in finance by far, followed by electronic technology, health technology, producer manufacturing, and retail trade. Several participants found that larger firms are adopting AI first. Using global AI training and skills data, Vinod Bakthavachalam distinguished cutting-edge, competitive, emerging, and lagging countries in terms of AI development.

- AI was viewed as having major implications for jobs, as it furthers automation. There was consensus that the nature of most jobs will change and that AI would place downward pressure on the wages of some types of jobs and upwards pressure on the wages of AI-related jobs. The rise in inequality is mainly driven by job polarization, a tendency that can be observed already prior to the rise of the current AI boom and has affected countries at all income levels in the past, including emerging and low-income countries. 
 
- Participants viewed enabling policy and regulatory environments for AI as critical to benefit from AI and minimize its risks. The group discussed the need to create virtuous circles between widely adopted platforms that ensure interoperability among AI systems; access to skills and human capital, access to data, software and hardware resources needed for AI. Karine Perset, Ilana Golbin, and Anand Rao explained their respective efforts to track national governments’ AI policy initiatives to understand government concerns and initiatives and to identify good practices. Charina Choi provided an overview of principles for public-private partnerships (PPPs) in AI development. Alessandro Annoni underlined the need for policies and regulations to build trust in AI and in AI companies.

- The group also discussed potential risks of AI on increasing inequality and polarization within society. The lack of good indicators on AI for developing countries remains an impediment. Some felt that those displaced by AI might need help during a transition phase, such as data entry clerks who may not have the skills to fill new jobs created by AI applications in areas like telemedicine. Ekkehard Ernst reported that AI was increasing inequality and job polarization worldwide, rather than increasing unemployment, with real compensation per worker increasing since 2001 in only the top 5\% “frontier firms”. Another concern was the AI brain drain from academia to large private sector companies, articulated notably by Zhao Jin who looked at the movement of AI professors/faculty in the US. Jin said that this phenomenon --- exponential since 2010 --- would reduce the number of future AI entrepreneurs.

- Consistent ways to quantify investment in AI were also discussed. In similar manner to measurement in AI private equity investment, national statistics agencies methodologies could be updated to capture public investments in AI R\&D. Daria Mehra explained the development of an AI network map of over 4,400 AI companies and the distinction between patterns of AI investment between China and the US. Michael Page estimated US Government investment in AI using data on USG transactions, solicitations, and budgets and put forward that the role of the federal government in AI funding might be evolving given private sector support for AI research on the international level.

- The productivity paradox and the time lag between adopting AI and seeing productivity gains materialize were also central. Erik Brynjolfsson stressed that although there are significant intangible investments in AI, it takes a long time to adapt organizational processes and overcome resistance to change and thus to enjoy productivity gains. Ramesh Johari also highlighted the need --- in addition to data and skills --- to focus on organizational change as companies look ahead at how they will process their data over the coming few years. 

- Other topics related to the impact of AI on the economy, notably AI’s impact on the functioning of financial markets --- both contributing to financial instability on the one hand, and on the other, helping to detect macro and micro risks. Peter Sarlin explained how central banks are using AI for financial stability surveillance. Christos Makridis emphasized the importance of cybersecurity and free AI education resources such as Coursera.

\subsection{\textit{Summary of Sub-Group 3: AI for Sustainable Development and Human Rights: Inclusion, Diversity, Human Dignity}}

This session focused on the measurement challenges related to AI applications of sustainable development and AI risks. Some of the core problems discussed included:

- AI has the potential to contribute to addressing societal challenges spanning all 17 of the United Nations SDGs, but there are bottlenecks including structured data, talent, implementation challenges. Monique Tuin also highlighted risks including explainability of AI decisions, addressing bias, managing data privacy and security, and considering how AI could be used (or misused) by various actors.

- Due to the classified nature of government information related to the development of fully autonomous weapons, it is difficult to further research and provide guidance on informed public understanding of this important topic. Marta Kosmyna surveyed global public perception, expert curated data, and measurement challenges about autonomous weapons.

-  Only select aspects of machine learning are incentivized within the academic community, while other crucial stages of machine learning are neglected, including phrasing a problem as a machine learning task, collecting data, interpreting results, publicizing results to the relevant community, and persuading users to adopt a new technique. Rachel Thomas highlighted that metrics tend to focus on easy-to-measure values, as opposed to genuine impact; that we cannot know everything we will need to measure in advance; and that genuine inclusion of diverse stakeholders is difficult to measure.

- Capacity, enabling environment, and accountability are critical to the adoption of AI in economic development, especially across Sub-Saharan Africa. We have to deal with the low availability of academic programs and their enrollment in Africa, an enabling environment for innovation, public and private investment to fuel the growth of AI and algorithmic transparency. Muchiri Nyaggah presented large AI initiatives currently underway in Sub-Saharan Africa. 

- Results from eight experiments suggest that laypeople show “algorithm appreciation” if people are easily susceptible to algorithmic advice. Don Moore discussed how we design systems to accommodate potential cognitive risks at individual and community levels that bring visibility and transparency to how human beings make decisions augmented by AI-based salient information.

- Women were underrepresented, making up just 18\% of the researchers publishing at AI conferences. Yoan Mantha discussed the importance of more rigorous data to classify AI and gender participation to provide sound policy guidance on pathways for women in AI at work. 

- AI talent pool is highly mobile, with about one-third of researchers working for an employer based in a country that was different from the country where they received their PhD. Deeper research is required on immigration policy and trade in services policy at the multilateral level to balance the skill-flow of AI among countries for competitiveness.

- Risks and threats of AI, including surveillance, privacy concerns, fake news, Twitter bots, and deep fakes are hard to measure at a local or community level. More fundamentally, discussions questioned what kind of world we want to create and how AI can help us get there. Many moderators and panelists including Eileen Donahoe, Margarita Quihuis, Jamila Loud, Anand Rao, Timnit Gebru discussed questions on how AI and AI research can be in the service of human rights and how do we keep human rights from not being an afterthought in AI development and research. Participants raised critical challenges, such as climate change and pandemics, and discussed how the community can be mobilized to help solve them. 

\section{Summary of All Talks}
\label{sec:headings2}

In this section, we provide a detailed summary for each individual talk from each breakout session. Section 3.1 covers talks on research and development. Section 3.2 covers talks on economic impact and societal considerations for policy decisions. Section 3.3 covers AI for sustainable development and human rights.

\subsection{Research and Development (including Technical Performance)}
\label{sec:headings3}


\subsubsection[Technology Vectors for Intelligent Systems Ashley Llorens, Applied Physics Lab (APL), Johns Hopkins University]{\texorpdfstring{\textit{Technology Vectors for Intelligent Systems} \\ Ashley Llorens, Applied Physics Lab (APL), Johns Hopkins University}{Technology Vectors for Intelligent Systems Ashley Llorens, Applied Physics Lab (APL), Johns Hopkins University}}

This presentation focused on long-term issues of evaluating the capabilities and robustness of systems in areas like space exploration and robotic prostheses. It derived four "technology vectors" that the AI community could attempt to quantify and measure to assess progress towards building systems that integrate advances in AI, robotics, and other technologies. These vectors are:

- \textbf{Autonomous perception:} systems that reason about their environment, focus on the mission-critical aspects of the scene, understand the intent of humans and other machines, and learn through exploration.

- \textbf{Superhuman decision-making and autonomous action:} systems that identify, evaluate, select, and execute effective courses of action with superhuman speed and accuracy for real-world challenges.

- \textbf{Human-machine teaming at the speed of thought:} systems that understand human intent and work in collaboration with humans to perform tasks that are difficult or impossible for humans to carry out with speed and accuracy.

- \textbf{Safe and assured operation:} systems that are robust to real-world perturbation and resilient to adversarial attacks, while pursuing  goals that are guaranteed to remain aligned with human intent.

\paragraph{What should researchers do?}
Researchers should try to develop metrics that capture progress along these four vectors. Measurements that work for integrated systems like this will facilitate the development of real-world sophisticated systems that continuously team with people.


\subsubsection[Towards More Meaningful Evaluations in AI Christopher Potts, Stanford University]{\texorpdfstring{\textit{Towards More Meaningful Evaluations in AI} \\ Christopher Potts, Stanford University}{Towards More Meaningful Evaluations in AI Christopher Potts, Stanford University}}

This presentation focused on the drawbacks of today's methods for evaluating technical progress, especially within NLP. Today, many researchers claim "human-level performance" on language tasks. These claims are often mis-construed as being about general human capabilities (e.g., question answering), but they are always about achieving certain scores on narrowly defined benchmarks. Recent developments in adversarial evaluation suggest ways to evaluate systems more robustly, highlighting their narrowness. 

\paragraph{Adversarial Evaluation}
\hfill 

Adversarial Evaluation is about developing evaluation methods that are hard for today's machines but would seem easy or trivial for humans. For instance, contemporary NLP systems that obtain high-scores on existing benchmarks for predicting the next words in sentences can be attacked with examples that a human would get right, e.g., if you feed the sentence "Father Chrostmas is also known as" to an AI system and ask it to predict the next two words, there's a reasonable chance it will get this wrong, whereas a human will spot the incorrect spelling and correctly guess "Santa Claus". Adversarial evaluation can take a variety of forms, for instance, by changing synonyms to antonyms in language tests and seeing how systems do.

\paragraph{What should researchers do?}
Researchers should try to create more adversarial evaluation approaches, so they can better understand technical progress (and brittleness) in these domains. They should also try and ask questions to motivate work. Three useful questions seem to be:

- Can a system behave \emph{systematically}, for example, making the same prediction (even if inaccurate) across examples that have been changed in irrelevant ways?

- Can a system assess its own confidence --- know when \emph{not} to make a prediction?

- Can a system make people happier and more productive in a specific task they perform --- and can we measure this?


\subsubsection[Data-Informed AI Policy Analysis Dewey Murdick and Michael Page, Center for Security and Emerging Technology (CSET), Georgetown University]{\texorpdfstring{\textit{Data-Informed AI Policy Analysis} \\ Dewey Murdick and Michael Page, Center for Security and Emerging Technology (CSET), Georgetown University}{Data-Informed AI Policy Analysis Dewey Murdick and Michael Page, Center for Security and Emerging Technology (CSET), Georgetown University}}

This presentation focused on using a combination of analytic and data-informed techniques to answer questions relevant to policy making at the intersection of AI and national security. The approach is conducive to addressing a wide variety of questions by connecting policy-aware analysis with a wide variety of datasets, such as scholarly literature, dissertations and theses, technical news, grant solicitations and awards, financial transactions for publicly and privately held corporations, venture capital financial transactions, government solicitations and procurement transactions, job postings, and career histories.
 
Multiple types of questions were briefly explored in the presentation. One example question was “Will U.S.-based big tech companies dominate the frontier of AI R\&D in 3-5 years?” This question was analyzed from the following angles over time:

- Comparing industry vs. academic research output by percentage of top papers and corporate research participation;

- Estimating talent-hiring rates by exploring the absolute and relative number of job postings by AI-relevant skill sets;

- Monitoring investment and funding trends from publicly observable grants, contracts, and public and private company investments; and

- Measuring research community growth rates and organizational membership to estimate when they will reach a relevant level of maturity.

\paragraph{What should researchers do?}
Researchers should make sure that their measures and metrics are contextually linked to relevant policy questions. This helps make sure that confusion about why they matter or how they should be interpreted can be minimized. Additionally, researchers should continue to explore ways to link datasets (e.g., by organizational entity name) to unlock additional insights that can only be discovered across multiple datasets.


\subsubsection[MAG: Leveraging AI for Scholarly Knowledge Acquisition and Reasoning Kuansan Wang, Yuxiao Dong, Zhihong Shen, Microsoft Research]{\texorpdfstring{\textit{MAG: Leveraging AI for Scholarly Knowledge Acquisition and Reasoning} \\ Kuansan Wang, Yuxiao Dong, Zhihong Shen, Microsoft Research}{MAG: Leveraging AI for Scholarly Knowledge Acquisition and Reasoning Kuansan Wang, Yuxiao Dong, Zhihong Shen, Microsoft Research}}

This presentation focused on a project that was initiated inside Microsoft Research to explore the extent to which modern AI technologies, particularly in the areas of natural language understanding and machine cognition, can be harnessed to assist researchers and technologists to keep up with the massive amount of literature and maintain or even improve their productivity. A major outcome of the project is Microsoft Academic Graph (\href{https://academic.microsoft.com/home}{MAG}) \cite{wang2019opportunities}. MAG is a knowledge graph that captures what concepts have been published in which articles. In contrast to similar datasets from commercial entities, MAG is largely curated by machines that automatically extracts knowledge from the entire web document collection indexed by Bing. 

MAG inherits the scale of an industrial-grade web search engine and is indeed recognized in the academic literature as providing more comprehensive and fresh coverage in more fields with high accuracy \cite{sinha2015overview}. Additionally, rooting deeply into the web search technology enables MAG to adapt numerous unsupervised machine learning techniques against malicious web contents for the detection of questionable citation behaviors and the so-called predatory publishers in scholarly communications \cite{wang2020microsoft}. This detection is feasible because predators appear like web spams and link farms and the objective is to depress their apparent importance despite their quantities.

\paragraph{What should researchers do?}

The result is a measure called saliency that, in the same manner as the PageRank of a web page, quantifies the probability of any MAG node being universally recognized as important. The efficacy of saliency is an ongoing research topic, although early evidence suggests it possesses many desired properties without the known drawbacks of frequently used metrics such as the citation count, h-index, and publication count in high impact factor venues \cite{wang2019review}. Aside from fully disclosing the methodologies, the biweekly updated MAG is also made freely available so that all the results are reproducible and all potential biases in the datasets are patently visible. Although it takes a considerable amount of extra efforts, open data and transparent methodologies are a necessity towards the ethical use and responsible development of AI.


\subsubsection[Peer-reviewed research --- volume and quality metrics Maria de Kleijn-Lloyd, Elsevier]{\texorpdfstring{\textit{Peer-reviewed research --- volume and quality metrics} \\ Maria de Kleijn-Lloyd, Elsevier}{Peer-reviewed research --- volume and quality metrics Maria de Kleijn-Lloyd, Elsevier}}

This presentation focused on “using AI to define AI” by mining keywords from various sources and then writing a classifier to identify AI research. It showed how Elsevier is using automated methods to automatically analyze the AI publication landscape, and cluster AI research into sub-themes for deeper trend analysis. 

It takes a well-structured database like Scopus --- linking articles to authors and institutions --- to get insights beyond simple volume metrics. Key analyses that can be supported are:

- Article (including conference proceedings and reviews) volumes per sub-theme per geography

- Identifying key research institutes that have both high volume and high citation impact, including corporate actors

- Brain circulation: movements of researchers among geographies and between academic and corporate institutions; comparing productivity and citation impact of mobile vs. sedentary researchers

- Diffusion: how AI research is becoming an integral part of research topic clusters outside its traditional home in Computer Science

Scientometric analysis like the above provides crucial insight to support policy questions for universities and governments.

\paragraph{What should researchers do?}
Manually annotated examples of AI papers could be critical in developing guidelines for labels to track fields within AI specializations. More research is needed to help clearly approximate the boundaries of AI publication subjects, like NLP or computer vision.


\subsubsection[arXlive: Real-time monitoring of research activity in arXiv Juan Mateos-Garcia, Joel Klinger, Konstantinos Stathoulopoulos and Russel Winch, NESTA]{\texorpdfstring{\textit{arXlive: Real-time monitoring of research activity in arXiv} \\ Juan Mateos-Garcia, Joel Klinger, Konstantinos Stathoulopoulos and Russel Winch, NESTA}{arXlive: Real-time monitoring of research activity in arXiv Juan Mateos-Garcia, Joel Klinger, Konstantinos Stathoulopoulos and Russel Winch, NESTA}}

This presentation discussed arXlive (\url{https://arxlive.org}), an open-source data collection, enrichment, and analysis system for real-time monitoring of research activity in arXiv preprints. In detail, the authors’ collected all the publications on arXiv and linked them to the Microsoft Academic Graph (MAG), that has more than 230M publications, then they geocoded the authors’ affiliations by linking them to the GRID, an open database with information on research institutions. The presentation also showed novel metrics for publications on arXiv by estimating how dissimilar the TF-IDF vector of an abstract is from its 1,000 similar abstracts. 

The authors’ built a platform that provides three ways for users to interact with the arXiv data: 

- Enabling rich and comprehensive searches powered by a fast, query expansion approach that returns results containing not only the initial search term but also similar ones to it. 

- Assisting users to expand their technical vocabulary and identify specialized terms by navigating the search query space. All of the terms shown to the users are contained in the publication abstracts so they can be used directly in the search engine. 

- Updating the figures of \href{https://arxiv.org/pdf/1808.06355.pdf}{Deep learning, deep change? Mapping the development of the Artificial Intelligence General Purpose Technology} \cite{klinger2018deep} paper on a daily basis, providing policymakers and researchers access to the most recent results.

\paragraph{What should researchers do?}
There are many new measures and metrics that can be derived on top of the rich data and services aggregated by arXlive --- for  example, the authors are working on estimating the amount of funding going into AI research by parsing the paper acknowledgments from the full-text of the publications. Other aspects include improving the search results by adding a semantic search engine that allows long-text queries while we are working on providing daily updates of the figures of the paper on \href{https://media.nesta.org.uk/documents/Gender_Diversity_in_AI_Research.pdf}{Gender Diversity in AI Research} \cite{stathoulopoulos2019gender}. Lastly, authors’ are developing an interface to enable the visual exploration of the search space.


\subsubsection[Tracking Performance Improvement in Computer Vision Bernard Ghanem, King Abdullah University of Science and Technology]{\texorpdfstring{\textit{Tracking Performance Improvement in Computer Vision} \\ Bernard Ghanem, King Abdullah University of Science and Technology}{Tracking Performance Improvement in Computer Vision Bernard Ghanem, King Abdullah University of Science and Technology}}

The presentation focused on tracking progress in activity understanding, a sub-field of AI research that aims to develop systems that can parse human activity in videos and figure out what behaviors are taking place (e.g., running, or dancing). The two main tasks discussed were: 

- Activity classification in video: The datasets include YouTube8M \cite{abu2016youtube}, \href{https://deepmind.com/research/open-source/kinetics}{Kinetics} \cite{li2020ava, carreira2018short, carreira2019short}, \href{https://research.google.com/ava/}{AVA} \cite{li2020ava}. The the goal is to automatically predict what activity is likely occurring in a video.

- Activity detection in video: The most popular dataset is \href{http://www.activity-net.org/}{ActivityNet} \cite{caba2015activitynet, ghanem2017activitynet, ghanem2018activitynet}. The goal is to automatically predict when a particular activity is likely occurring in a video.

One of the challenges highlighted was that activity understanding and activity recognition research are relatively expensive tasks in terms of the data and compute requirements, which restricts the number of people that can do research in these areas, skewing our sense of progress.

\paragraph{What should researchers do?}

After four years of running the ActivityNet challenge, the presenter identified two main challenges in the ongoing measurement and assessment of activities in videos: 

- Though the performance in each task is improving from year to year, the rate of this improvement is slowing down. This limitation suggests we need new research approaches to create more efficient methods. The performance on these tasks remains far from human performance.

- Current state-of-the-art methods for activity classification and detection do not seem to exploit temporal or spatio-temporal context well enough to accurately classify and localize an activity in video. Specifically, long activities with unique/distinctive patterns in motion cues, objects, and human-object interactions are the easiest to classify and detect (e.g. zumba and rock climbing); while shorter activities containing more subtle motion or object patterns are much harder (e.g. drinking coffee and smoking). This indicates a need for new generation methods that exploit semantic context to better classify and detect activities in video.


\subsubsection[Recent Advances in Natural Language Inference: A Survey of Benchmarks Resources, and Approaches, Shane Storks, Qiaozi Gao, Joyce Y. Chai, University of Michigan, Ann Arbor, and Michigan State University]{\texorpdfstring{\textit{Recent Advances in Natural Language Inference: A Survey of Benchmarks} \\ Resources, and Approaches, Shane Storks, Qiaozi Gao, Joyce Y. Chai, University of Michigan, Ann Arbor, and Michigan State University}{Recent Advances in Natural Language Inference: A Survey of Benchmarks Resources, and Approaches, Shane Storks, Qiaozi Gao, Joyce Y. Chai, University of Michigan, Ann Arbor, and Michigan State University}}

This presentation focused on NLP research into deep language understanding beyond explicitly stated text, relying on inference and knowledge of the world. Many benchmark datasets and tasks have been created to support the development and evaluation of such natural language inference ability. To facilitate quantitative evaluation and encourage broader participation, various leaderboards are consolidated in an arXiv paper by the authors \cite{storks2019recent}. The authors’ provided an overview of existing benchmark datasets widely used in the Natural Language Inference (NLI) community. 

The authors  noted that we need stronger justification and a better understanding of design choices for models. For example, design choices like parameter tuning strategies are often overlooked in favor of more significant or interesting model improvements, but these small choices can actually have a significant effect on performance. More complex models may lead to better performance on a particular benchmark, but simpler models with better parameter tuning may later lead to comparable results. A related topic is domain adaptation, i.e., where the distribution of training data differs significantly from the distribution of test data. This is a good step toward generalization to unseen data. State-of-the-art models like BERT are getting somewhat closer to this capability, being pre-trained on large text corpora and capable of being fine-tuned to new problems with minimal effort or task-specific modifications. 

\paragraph{What should researchers do?}

Researchers should build benchmarks that will help us develop more advanced natural language systems. The presenters suggested three promising avenues to explore: 

- Develop benchmarks with a greater emphasis on external knowledge acquisition and incorporation; It may be worthwhile to explore new task formulations beyond the text that involve artificial agents (in either a simulated world or the real physical world) which can use language to communicate, to perceive, and to act including interactive task learning, implementing a physically embodied Turing Test, humans exploring limitations in understanding language based on interaction with AI systems, and so on).

- Design benchmarks that more directly evaluate the reasoning capabilities of models. Recent results have questioned whether state-of-the-art approaches actually perform genuine inference and reasoning for those benchmark tasks. NLI systems can break down due to small, inconsequential changes in inputs. It is important for methods to automatically integrate many types of reasoning such as temporal reasoning, plausible reasoning, and analogy. The majority of benchmarks do not support a systematic evaluation of reasoning abilities. 

- Special efforts should be made to reduce/eliminate superficial data bias in order for these benchmarks to be useful.


\subsubsection[Measuring AI in Defense Applications: The Need for Safety and and Reliability Measures Andrew J Lohn, RAND Corporation]{\texorpdfstring{\textit{Measuring AI in Defense Applications: The Need for Safety and and Reliability Measures} \\ Andrew J Lohn, RAND Corporation}{Measuring AI in Defense Applications: The Need for Safety and and Reliability Measures Andrew J Lohn, RAND Corporation}}

This presentation focused on the measurement challenges of the Department of Defense (DoD) and military commanders. High-risk defense AI applications present a new set of measurement challenges, including: 

- Safety and Reliability metrics for defense-related AI systems: Defense acquisition mandates more stringent safety requirements than commercial systems, so new measurements are required to assure performance in these scenarios. Some particularly important are measuring system performance in response to an adversary who may seek to exploit vulnerabilities in the system's algorithms, implementations, or data.

- Generalization: Defense applications must be able to respond appropriately given environments and inputs that were not explicitly foreseen. AI may need to be tested in the way that other intelligences are: through licensure. This effort will require the ability to determine which tasks are different enough to require separate licensing, how much can a system learn or can its environment evolve before it needs to be relicensed, and how diverse a range of implementations can be covered by a given licensing process.

\paragraph{What should researchers do?}

Metrics and evaluations do not yet exist  in sufficiently mature forms for it to be easy to license AI technology into defense. We need to develop safety metrics and tests, along with suites of tests. These suites will need to  address questions such as which tasks are different enough to require separate licensing, and how diverse a range of implementations can be covered by a given licensing procedure. Licensure is one of many approaches to increase the safety and reliability of AI systems in general, particularly for military systems where consequences can be particularly dire.


\subsubsection[Keeping up with ML progress using Papers with Code Robert Stojnic, Papers with Code]{\texorpdfstring{\textit{Keeping up with ML progress using Papers with Code} \\ Robert Stojnic, Papers with Code}{Keeping up with ML progress using Papers with Code Robert Stojnic, Papers with Code}}

This presentation focused on \href{https://paperswithcode.com/}{Papers with Code} (PwC) --- a free and open-source resource that puts together machine learning papers, code, and evaluation tables on the web. It is built together with the community and powered by automation to help track the state of the art performance (SOTA) for specific datasets and tasks by extracting performance results in papers and open-source code. At the time of writing, PwC contains:

- More than 120,000 papers

- 23,000 papers with code implementations 

- More than 2,500 leaderboards across all areas of ML, including NLP, CV, Speech, Reinforcement Learning, and many more. 

This free resource has been built using a combination of automation and community work; papers and code repositories are automatically gathered and connected, and leaderboards are automatically imported. 

\paragraph{What should researchers do?}
Currently, Paperswithcode leaderboards are manually annotated tasks and datasets with some automation. Could we build a fully autonomous agent to curate, validate, and track technical progress over time by reading full-text papers and benchmark SOTA performance for specific tasks at scale?

\subsubsection[Evaluation in Commercial Machine Translation Systems Konstantin Savenkov and Grigory Sapunov, Intento]{\texorpdfstring{\textit{Evaluation in Commercial Machine Translation Systems} \\ Konstantin Savenkov and Grigory Sapunov, Intento}{Evaluation in Commercial Machine Translation Systems Konstantin Savenkov and Grigory Sapunov, Intento}}

The number of commercially available MT systems has grown from 13 in March 2018 to 26 in November 2019, according to an analysis by Intento. The main factor here is the availability of open-source Neural machine translation (NMT) frameworks, which enable regional data owners to invest in the data curation and roll-out of niche MT engines which are competitive with offers from the global players. Most commercial MT systems are built in Europe (9), then China (8), and the US (5). In terms of the language pair support, 5 global systems support more than 3,000 language pairs, 7 in the middle tier with 300-1000 language pairs, and the rest  support 20-100 language pairs (slide 23 in \href{https://www.slideshare.net/KonstantinSavenkov/state-of-the-machine-translation-by-intento-stock-engines-jun-2019}{presentation}). 

Another important driver in commercial MT is the advent of domain-adaptive MT, which allows creating specialized MT models by adapting a general-purpose model with a (comparably) modest amount of training data. These models are hosted in the cloud using the provider infrastructure, significantly lowering the required level of ML or engineering expertise on the client side. 
 
The main findings are:

- The performance of the baseline models varies a lot between language pairs. Commercial MT quality is evaluated using the hLEPOR metric. hLEPOR score of 0.7 means almost human-level quality with just a few mistakes. The achievable hLEPOR score ranges from 0.3-0.4 for some language pairs (Chinese-Italian, French-Russian, Japanese-French) to 0.7-0.8 for others (English-German, English-Portuguese).

- The main driver of performance is language pair popularity, which defines how much investment goes into data acquisition and curation. Also, the next-generation translation technology (such as Transformer) is being rolled out to the most popular language pairs first, while rare language pairs may still employ Phrase-Based Machine Translation (PBMT) models or perform pivot translation through another language (usually, English).

- The wide range of performance across language pairs is not a mere snapshot but also demonstrates progress in the performance of commercial MT systems.

\paragraph{What should researchers do?}

Researchers can bring greater attention to low-resource (LR) language pairs and help make faster technical progress on LR language pairs. This is an area where there are fewer incentives for large commercial entities to develop high-performing systems. Another important field of work is reducing the appearance of gender bias in MT systems.


\subsubsection[On Quantifying the Role of Ethics in AI with Applications on the Assessment of Gender Bias in Machine Translation Pedro H.C. Avelar, Marcelo Prates, Luis C. Lamb, Universidade Federal do Rio Grande do Sul]{\texorpdfstring{\textit{On Quantifying the Role of Ethics in AI with Applications on the Assessment of Gender Bias in Machine Translation} \\ Pedro H.C. Avelar, Marcelo Prates, Luis C. Lamb, Universidade Federal do Rio Grande do Sul}{On Quantifying the Role of Ethics in AI with Applications on the Assessment of Gender Bias in Machine Translation Pedro H.C. Avelar, Marcelo Prates, Luis C. Lamb, Universidade Federal do Rio Grande do Sul}}
This presentation focused on two approaches to quantify ethics in AI: the first on measuring gender bias in Machine Translation and the second on measuring whether ethics itself is discussed in papers published in the main tracks of flagship AI and robotics venues, building upon \cite{PratesAL18}. 

Neural machine translation (NMT) is becoming ubiquitous, and since such tools achieve human parity in some cases, the potential bias of such systems, caused by imbalances in training sets, is a widespread concern \cite{PAL2019}. The study compared the distribution of translations from gender-neutral languages into English, with gender labor force participation information from the United States Bureau of Labor Statistics.

On quantifying the role of ethics in AI research, this work looked for keywords in paper titles and abstracts published in flagship AI and robotics venues, and by comparing the number of  ethics-related keywords with a control set of “classical” and “trending” AI keywords.

\paragraph{What should researchers do?}
Four actions are proposed to alleviate gender bias in NMT:

- MT tools could provide confidence scores for individual words, allowing users to identify how severe is the bias present by the tools 

- One could use a test suite to evaluate the system for biases, much like the one used in this case study, and either report such biases to its users or decide not to release the tool

- A subset of the training dataset could be curated to mitigate biases learned in other portions of the data.

- Regarding ethics at AI conferences work, researchers could explore a more comprehensive taxonomy of ethics-related keywords for bibliometrics analysis.


\subsubsection[Quantifying Algorithmic Improvements over Time Lars Kotthoff, University of Wyoming]{\texorpdfstring{\textit{Quantifying Algorithmic Improvements over Time} \\ Lars Kotthoff, University of Wyoming}{Quantifying Algorithmic Improvements over Time Lars Kotthoff, University of Wyoming}}

This presentation focused on quantifying the improvement that a particular contribution has made to the state of the art. Generally, competitions almost exclusively focus on standalone performance --- each submission is run on a set of benchmark problems, and the submission with the overall best performance wins. The state of the art in this field is now defined by the winner of the competition, even though it may not achieve the best performance on all benchmark problems. Arguably, the other submissions also contribute to the state of the art --- they might focus on narrower parts of the benchmark space and show overall worse performance, but really excel on what they focus on. This is completely ignored by standalone performance. 

The author proposed to use the Shapley value, a concept from co-operative game-theory that assigns a unique distribution of total surplus generated by a coalition of players. In this instance, the Shapley value was used to assess the contributions of individual submissions to a competition to the overall state of the art \cite{frechette2016using}. Importantly, it does not suffer from the problem outlined above and gives credit to all submissions that show promise on any part of the benchmark set. The proposed approach enjoys all the guarantees of the Shapley value, namely additivity (allowing multiple performance measures to be combined), efficiency (the overall value is distributed entirely), dummy player (a contribution that does nothing gets no credit), and symmetry (two submissions that are identical have identical contributions).

The author proposed a temporal modification of the Shapley value that takes into account that the state of the art in a field was not defined in one fell swoop, but developed over time \cite{kotthoff2018quantifying}. Later approaches are based on earlier ones that should receive credit for paving the way --- the temporal Shapley value achieves this by allowing submissions to contribute only to the state of the art after they were created, while retaining the desirable properties of the Shapley value. 

\paragraph{What should researchers do?}

Researchers should use the Shapley value to evaluate competitions and the temporal Shapley value to track the progress of the state of the art in a field over time. Using these measures instead of the standard standalone performance gives a more accurate picture of the state of the art and encourages collaboration to advance a field as a whole rather than incentivizing small improvements across large and diverse sets of benchmarks. Research efforts can be focused on particular sub-fields while receiving credit for such focused advances.


\subsubsection[Human Forecasts for Technical Performance Progress Ben Goldhaber, Metaculus]{\texorpdfstring{\textit{Human Forecasts for Technical Performance Progress} \\ Ben Goldhaber, Metaculus}{Human Forecasts for Technical Performance Progress Ben Goldhaber, Metaculus}}

This presentation focused on calibrated forecasts from human beings to guide strategic decisions within the AI field, for example, AI technical benchmarks and landmark events. Between Dec 2018 and Dec 2019, \href{https://www.metaculus.com/}{Metaculus} (\url{ai.metaculus.com}) and AI Index conducted a prediction tournament on questions relevant to the AI research community.

Through a series of structured interviews, 100 questions expected to be valuable to policy makers were solicited and written. Then, using Metaculus as a platform, a community of forecasters and researchers was asked to register predictions on the questions. In total over the past year, 2700 predictions were made. While many of the questions have not yet been resolved, to date the community has an average Brier Score (proper score function that measures the accuracy of probabilistic predictions) of 0.146, notably better than an uninformed forecaster.

\paragraph{What should researchers do?}
How can we extend human forecasts to cover more questions of interest and incorporate structured clusters of questions? Researchers should develop key questions for human forecasts on AI's technical performance and broader implications for curated expert foresights.

\subsection{Economic Impact and Societal Considerations for Policy Decisions}
\label{sec:others1}


\subsubsection[China AI Index: China-US Comparison Chenggang Xu, Cheung Kong Graduate School of Business (CKGSB) and London School of Economics (LSE)]{\texorpdfstring{\textit{China AI Index: China-US Comparison} \\ Chenggang Xu, Cheung Kong Graduate School of Business (CKGSB) and London School of Economics (LSE)}{China AI Index: China-US Comparison Chenggang Xu, Cheung Kong Graduate School of Business (CKGSB) and London School of Economics (LSE)}}
The presentation focused on China-US comparisons from the \href{https://english.ckgsb.edu.cn/wp-content/uploads/2019/11/\%E4\%B8\%AD\%E5\%9B\%BD\%E4\%BA\%BA\%E5\%B7\%A5\%E6\%99\%BA\%E8\%83\%BD\%E6\%8C\%87\%E6\%95\%B0-final.pdf} {China AI Index}. One of the major purposes of this project is to improve our understanding of the upcoming industrial revolution, driven by AI, in different economic systems. The failure of the Soviet Union in semiconductors, ICs, and computers were deeply rooted in its system \cite{qian1998innovation, xu2017capitalism}. This project is designed to help  reveal the performance of the Chinese system, which is a reformed Soviet system, vis-à-vis that of the counterpart in the US. The following evidence presented was revealing on the US-China economic dynamics:

-  US AI firms are either very large or small, whereas most AI firms in China are medium-sized. A typical US AI firm has no more than 10 or a few dozen employees. The United States also has five large AI companies that employ more than 5,000 workers while China has none.     

- The gap between China and the US is shrinking in certain areas such as computer vision research or broader AI publications. In other areas such as basic research or papers with at least 1,000 citations, there is no evidence of convergence. For example, the number of Chinese and US CVPR conference participants converged by 2018 and in total publication and total citation, the gaps between the US and China continue shrinking. However, in areas that are more in basic research, there is no observable convergence. For example, mostly US authors are participating in ML Conference or papers with more than 1,000 citations. 

- Chinese researchers rely heavily on US open-source software packages for basic algorithms based on data from Github. Most of the open-source AI software packages used (starred) by Chinese AI researchers are developed by US institutions, and the most popular one is TensorFlow developed by Google. In general, the number of US-developed open source AI software packages starred by researchers from both countries is twenty times more than those developed by Chinese institutions.

\paragraph{What should researchers do?}
Researchers should study the role of institutions and market structure on the success and failure of AI projects between two distinct market designs. The more efficient R\&D system must try a large number of projects and rely on the market to screen ex-post. But this mechanism can work only when the system is able to commit to terminate unpromising projects. When a system is not able to do so, they will replace the ex-post market screening mechanism, by ex-ante bureaucratic selection. In this case, the number of projects will be much smaller and each ex-ante selected project will receive more resources. The author proposed that  given the uncertain nature of the technology, it is likely that ex-ante betting on small number projects will be out-performed by the system which is able to bet on a large number of projects. Researchers should conduct theoretical and empirical research to improve our understanding of the relationship between market design and the successful execution of AI projects.  


\subsubsection[Measuring AI investment: Why does it matter and why is it challenging? Prasanna Tambe, Wharton School, University of Pennsylvania]{\texorpdfstring{\textit{Measuring AI investment: Why does it matter and why is it challenging?} \\ Prasanna Tambe, Wharton School, University of Pennsylvania}{Measuring AI investment: Why does it matter and why is it challenging? Prasanna Tambe, Wharton School, University of Pennsylvania}}
This presentation focused on two areas of measurement in AI: its use and its impact. It highlighted that measuring AI's value may be harder than it has been for its technological precursors. Extensive regulatory attention is being focused on AI developments, as well as the legal factors and new policy initiatives that govern the use of data and algorithms, such as the General Data Protection Regulation (GDPR), which can affect how AI technologies will be deployed. The returns to AI technologies and the pace at which these are realized, therefore, may be difficult to anticipate and profoundly different across nations, states, or even cities and counties. 

On measuring the use of AI:

A number of disparate inputs are required to put AI into production — human capital, software, data, computational power, and new management practices — none of these are easy to observe and statistical agencies do not regularly collect data on them. For mature information technologies such as network administration, database management, and web development, some of the inputs to technology investment (such as software licenses) are available and can be used to track usage effectively. However, firms are still adjusting inputs to AI and experimenting with the technology, so even investments that can be observed tend to be noisy indicators. For instance, rapid changes to AI technologies can change the value of investments in AI-related skills \cite{rock2019engineering}. Similarly, heavy reliance on open-source software in corporate data science makes any data that is available on AI software spending likely to be noisy. Other inputs that might indicate AI use, such as specific types of data assets or differences in firms’ management practices, are generally even more difficult to track consistently.

On producing robust evidence on the impact of AI:

The regulatory and environmental constraints, as well as “hidden” factors such as differences in the quality of firms’ data assets, can influence the returns to AI technologies. Moreover, many corporate AI projects are in the trial phase, and have yet to be deployed. Data on the use of AI within the firm, therefore, may not yet correlate with any measurable impact. Differences in management practices can also have an important role to play in gains from new technologies, but management practices are costly and time-consuming for firms to install or change. Indeed, there can be a significant time-lag between when new technologies such as AI are adopted by industries and when they start to produce a measurable difference in productivity.

\paragraph{What should researchers do?}
The availability of new and increasingly granular data will be valuable to better measure the use of AI and study the impact of AI including: 

- Data on corporate activity collected through digital channels is a formidable weapon for empirical researchers and has the potential to significantly improve our ability to track changes in firms’ technological configurations \cite{horton2015labor}. 

- Some of the author’s recent work has used data such as those collected and curated by platforms such as LinkedIn and Burning Glass Technologies to generate measures of investments in skills related to data and algorithmic decision-making \cite{tambe2014big, tambe2020paying,tambeetal2019}. 

- Software use might be tracked through other digital platforms, such as GitHub, in similar ways. Of course, it may be the case that the factors that these data sources cannot shed light on, such as firms’ management practices, might be the most important ones for explaining returns to AI technological progress.


\subsubsection[Measuring AI Dynamism and Impact on Inequality Guy Berger, LinkedIn]{\texorpdfstring{\textit{Measuring AI Dynamism and Impact on Inequality} \\ Guy Berger, LinkedIn}{Measuring AI Dynamism and Impact on Inequality Guy Berger, LinkedIn}}
This presentation focused on three areas of economic measurement:

- \textbf{The slowing dynamism conundrum:} A wide range of metrics --- productivity growth, business creation and destruction, interest rates, etc. indicate that the pace of change and transformation in the US and global economies has slowed down dramatically over the past 10-15 years.  A deeper understanding of why innovative technologies like AI are not showing up as meaningful in the economic data and catching inflection points early is essential.

- \textbf{Can we quantify how technology is diffusing across regions, occupations, and industries?} Is reskilling happening at an individual level, or only at an aggregate level?

- \textbf{Distributional outcomes and inequality:} As political and social events across the globe suggest,being able to identify \say{winners and losers,} and designing institutions and technological implementation in a way that equitably shares the gains from technology gains, is going to be essential.

\paragraph{What should researchers do?}
Researchers should comprehensively study the impact of emerging technologies on the economy and society. Let’s assume that despite taking longer than the hype suggests, AI eventually transforms our economy in a meaningful way. It's not the only force that’s going to do so. We will see the impact of other technologies; we are going to experience potentially significant climate change, massive fluctuations of the business cycle, and policy changes that will make a huge difference.  We need to properly contextualize the economic impact of AI relative to those other forces, and to understand how it either complements or substitutes for them.


\subsubsection[Macro-economic impact of AI Ekkerhard Ernst, International Labor Organization (ILO)]{\texorpdfstring{\textit{Macro-economic impact of AI} \\ Ekkerhard Ernst, International Labor Organization (ILO)}{Measuring AI Dynamism and Impact on Inequality Guy Berger, LinkedIn}}
This presentation focused on AI and its impact on jobs, productivity, and inequality from a global lens. The presentation showed that:

- Automation has a significantly stronger impact on employment in developing countries: The impact of robotization on employment is more than ten times as large in emerging economies than in advanced economies. The impact comes from both a direct effect of automation on low-skilled jobs and a reshoring effect \cite{carbonero2018robots}.

- The impact of AI on jobs might be more diverse. AI is a general-purpose technology that can have effects both on labor and  capital productivity. A particularly large impact might be expected through improvements in network management to enhance Total Factor Productivity (e.g. electricity, transportation, waste management, networks of global value chains, etc.). The benefits of these AI innovations might be higher in emerging countries as they typically suffer from the lack of or low-quality infrastructure.

- To measure the relative impact of capital- vs labor-saving impacts of technology, the authors’ used a new indicator by \cite{felten2018method} to measure the extent to which AI transforms different occupations in the US. Following \cite{fossen2019mapping} they apply this measure together with the risk of computerization developed by \cite{frey2017future} to 4-digit occupational categories in the US, Thailand, and South Africa.

The findings include:

- For the US: A polarization of jobs is visible with 9 of the 10 largest occupations characterised by the high risk of automation (labor-saving AI) and in almost half of all occupations characterized by transformational AI (capital-saving AI).

- For Thailand (and South Africa): The polarization is titled much more towards occupations characterized by a high risk of automation. Moreover, a large share of occupations is characterized by disruption AND transformation, i.e. machines are likely to replace human labor even more rapidly and thoroughly than through simple automation.

\paragraph{What should researchers do?}
Researchers should study the dynamics of AI and inequality more than job loss, at least in OECD countries. To better understand the society-wide implications of AI, proper indices are needed to monitor AI deployment. Simply counting the number of patents or measuring digital bandwidth remains insufficient. The presentation suggested  four dimensions that warrant closer analysis:

- Network applications, for instance for Smart City deployments would affect electricity, transportation, urban planning, or trading networks. This research requires a closer look at the digital divide between advanced and emerging economies. A simple look at smartphone penetration rates might bring some initial information on how these technologies can be applied in specific circumstances (Deloitte produces a Smart City index).

- The current analysis of the economics of AI lacks a clear understanding of the resource implications of AI. Most types of analysis implicitly conclude that AI could reach a stage where it has absolute advantages over human labor. This assumption neglects the fact that AI applications already consume significant amounts of energy that are likely to prevent a full deployment of automating AI. On the other hand, AI applications explicitly targeted at a reduction in energy consumption (e.g. electricity grid management) could contribute to help reduce carbon emissions. AI applications should be indexed according to their net energy consumption and carbon footprint.

- AI applications have very different effects depending on their sectors. The potential for sectoral productivity gains might, therefore, differ significantly. Rather than characterizing AI applications according to their automation potential generally, it might be more useful to develop sector-specific AI indices.


\subsubsection[Measuring the Economic Effects of Artificial Intelligence Daniel Rock, MIT]{\texorpdfstring{\textit{Measuring the Economic Effects of Artificial Intelligence} \\ Daniel Rock, MIT}{Measuring the Economic Effects of Artificial Intelligence Daniel Rock, MIT}}
This presentation focused on AI as a new variety of GPT by using novel firm-level data. \cite{rock2019engineering} develops an online resume-based measure of firm-level AI talent using LinkedIn profile data and shows:

- A taxonomy that classifies self-reported skills as having AI content. 

- Uses firm-level measures of AI skill stocks to estimate the impact of AI technologies on the market values of those firms. 

This paper finds that:

- The launch of Tensorflow, a deep learning software module open-sourced by Google, caused a roughly 4-7\% increase in the market value of AI-using firms. 

- The capital base of AI-using firms became more valuable as a result of complementary AI talent becoming more abundant. 

- Firms that are more exposed to ML technologies via the tasks performed by their workers have declined in value in recent years. 

\paragraph{What should researchers do?}

Researchers should explore the following areas to improve data and related uncertainties to track AI adoption for firms and regions: 

- Data uncertainties to assess complementarity and substitution effects of AI skills. Some economic actors, like workers with AI skills, are poised to benefit from technological advances. Others might be negatively impacted as their skills or business models are obsoleted by new developments. This balance of complementarity and substitution is partially measurable via skills measurement on platforms like LinkedIn. These online resumes are attractive both because they can identify the occupational composition of firms (and therefore aggregated exposures to new technologies like AI), and also because indices of the organizational capital and skills necessary to implement technology are recoverable from what workers claim to know. Of course, there are substantial limitations in terms of coverage over time, cross-sectionally by worker and/or company, and in the type of skills that are reported. 

- Address sampling bias in job datasets. Econometric techniques to adjust for sampling bias and normalization of these datasets should be a major prerequisite for any analysis using these sources. Done well, as with job posting data from Burning Glass Technologies in \cite{hershbein2018recessions}, these "digital breadcrumbs" can be especially revealing about changes in the overall economy. Additionally, the macroeconomic effects of new technologies can be better resolved following frameworks like that in \cite{brynjolfsson2018productivity}.  


\subsubsection[A short introduction to CSET and US government spending on AI Michael Page, Center for Security and Emerging Technology (CSET), Georgetown University]{\texorpdfstring{\textit{A short introduction to CSET and US government spending on AI} \\ Michael Page, Center for Security and Emerging Technology (CSET), Georgetown University}{A short introduction to CSET and US government spending on AI Michael Page, Center for Security and Emerging Technology (CSET), Georgetown University}}
This presentation focused on US government AI spending data to guide policy decisions. The US government makes available several different categories of spending data. Transactions --- which include grants, contracts, loans, and other financial assistance --- are available on \url{USAspending.gov}. USAspending in turn collects data from several sources, including the Federal Procurement Data System (FPDS) for prime contract data. Solicitation data is available on \url{SAM.gov} (formerly Federal Business Opportunities). Budget data is available on different agencies’ websites. For example, historic Department of Defense budget data is available on the Defense Technical Information Center’s website.

\paragraph{What should researchers do?}
Researchers should use multiple sources of data to cross-reference and to analyze government spending on AI. Measuring public AI investment in the US poses several challenges:

- Data quality. Data from USASpending.gov, for example, is often missing or mislabeled.

- What’s “AI?" Identifying what spending is “artificial intelligence” related can be challenging for several reasons, including (i) transactions often contain little descriptive information; and (ii) some programs might be rebranded with AI keywords, leading to an inflated estimate of AI-related growth.

- International comparisons. Because government spending serves different functions in different countries, direct comparisons can be misleading.

- What is the federal government’s role in AI funding? Given the research community’s private-sector support and international character, the role of the federal government might be evolving.


\subsubsection[Measuring Private Investment in AI Daria Mehra, Quid]{\texorpdfstring{\textit{Measuring Private Investment in AI} \\ Daria Mehra, Quid}{Measuring Private Investment in AI Daria Mehra, Quid}}

This presentation focused on assessing the global, regional, and sector-based activities of AI startups and investors. The author used Quid to search for all startups that received more than \$400,000 in investment during the last ten years. The results showed that: 

- Global AI investment has continued to increase by an average of 48\% each year since 2014. 

- The US has held a dominant lead in terms of the number of startups and overall investment in the industry.  A few exceptionally high investment deals with Chinese startups over the last year have begun to close the gap. 

- Comparing investment totals on a per capita basis, however, Israel takes the top spot, followed by Singapore.

To surface popular destinations for investment, the presentation identified unique sectors within the larger AI industry. 

- During the last year, companies working on autonomous vehicles received the largest share of investment, followed by MedTech (specifically for cancer drugs and therapy) and facial recognition for law enforcement/surveillance. The largest growth sectors for AI technology included robotic process automation and supply chain management. 

- While AI startups in the US and Europe shared some common sectors, including MedTech, Text Analytics, and Retail Tech, AI companies in China were heavily focused on Industrial Automation (specifically oil \& gas), Facial Recognition, and EdTech. India had far fewer startups than other countries surveyed, with a greater interest in Robotic Process Automation, Credit Cards/Lending, and Hospitality/Travel.

\paragraph{What should researchers do?}

Researchers should explore robust taxonomies and semi-auto labeling methods to classify an AI company. It is also important to distinguish AI companies producing new AI from those applying it. A third-class of companies are just posers and write AI in their company information. With these distinctions in mind, researchers should integrate and analyze multiple sources of information about a company including corporate filings, earnings calls, news articles, patents, blogs, social media, etc. to comprehensively identify core AI components of a company. There is no database of intangible products and services that companies are deploying and thus it is very difficult to track how many products and services in the market may have AI components. Deeper research is required to build companies’ profiles and pathways of AI adoption to distinguish large and small companies.


\subsubsection[Where to Now: measuring specialization, disruption, and diversity in AI R\&D Juan Mateos-Garcia, Joel Klinger, Konstantinos Stathoulopoulos and Russel Winch, NESTA]{\texorpdfstring{\textit{Where to Now: measuring specialization, disruption, and diversity in AI R\&D} \\ Juan Mateos-Garcia, Joel Klinger, Konstantinos Stathoulopoulos and Russel Winch, NESTA}{Where to Now: measuring specialization, disruption, and diversity in AI R\&D Juan Mateos-Garcia, Joel Klinger, Konstantinos Stathoulopoulos and Russel Winch, NESTA}}

This presentation focused on measuring deep learning specialization and its contribution to application domains. Analyzing textual descriptions of AI R\&D with topic modeling algorithms shows the growing use of deep learning on arXiv:

- The rise of China as one of the global leaders in deep learning research

- The importance of co-location between research and industrial capabilities for developing strong deep learning clusters. 

- Countries with lower levels of political and civil freedom (and in particular, China) are overrepresented in the facial recognition field. 

- Thematic disruption brought about by the arrival of the deep learning paradigm, with formerly dominant topics related to symbolic and statistical approaches to AI losing centrality in favor of deep learning.

- Academic institutions tend to be more theoretically diverse than corporate labs pursuing narrow research agendas. These findings highlight the important role of public interest research and funding to preserve diversity in AI research.

\paragraph{What should researchers do?}

Researchers should leverage open source data and code to help track AI activity in a more dynamic manner. To enable policymakers and practitioners to explore the rich and complex landscape of AI R\&D that is revealed, researchers should use novel data sources and methods to dig beneath the surface of highly aggregated, monolithic indices of AI activity.


\subsubsection[AI Brain Drain Michael Gofman and Zhao Jin, University of Rochester]{\texorpdfstring{\textit{AI Brain Drain} \\ Michael Gofman and Zhao Jin, University of Rochester}{AI Brain Drain Michael Gofman and Zhao Jin, University of Rochester}}

As AI becomes one of the most promising and disruptive technologies, the top tech firms are trying to corner the market for AI talent, especially AI professors. AI professors at North American universities are being offered compensation packages by corporations that universities cannot match. This presentation presented the following new insights about the AI brain drain:

- Between 2004-2018, 221 professors at North American universities accepted an industry job. The reallocation of AI human capital from research universities to the private sector has important consequences for the entrepreneurship activity of the students in the affected universities.

- Following AI faculty departures, the authors’ found a negative causal effect on the quantity and quality of innovation, measured using the entrepreneurship data of 3,000 STEM alumni, 363 AI entrepreneurs, and 177 AI startups. They conclude that knowledge transfer via higher education is an important channel for the diffusion of AI-driven technological change. 

- AI faculty departures have a negative effect on the early-stage funding of AI startups with founders who graduated in year t from the affected universities. Specifically, a one standard deviation increase in the tenured professors' departures in time window [t-6, t-4] decreases, on average, first round and series-A round funding by, respectively, \$0.6 million and \$3.15 million. Relative to the sample average, these numbers imply a 22\% decrease in first-round funding and a 28\% decline in series-A-round funding. Moreover, a one standard deviation increase in the tenured professors' departures in time window [t-6, t-4] decreases funding growth from the first round to the second round by 20\%. 

- Departures by tenured professors' 4-6 years prior to the students' graduation year have the largest effect. Specifically, for a given university, a one standard deviation increase in tenured professors' departures during time window [t-6, t-4] on average reduces the number of future AI entrepreneurs who graduate in year t by 13\%. The effect is most pronounced for the top 10 universities, entrepreneurs with PhD degrees, and for startups in the field of deep learning.

\paragraph{What should researchers do?}
The AI brain drain from universities affects innovation by firms, universities, and students. It could be interesting to study the effect of AI faculty departures on the productivity of companies they join, the knowledge creation in the universities that they leave, and on the labor market outcomes of AI students in the affected universities. This analysis would give policy makers a more comprehensive picture of the net effects of the brain drain on productivity and externalities in the labor market for AI talent.


\subsubsection[Central Bank and Corporate Perceptions of Artificial Intelligence (AI) Evan A. Schnidman, Prattle]{\texorpdfstring{\textit{Central Bank and Corporate Perceptions of Artificial Intelligence (AI)} \\ Evan A. Schnidman, Prattle}{Central Bank and Corporate Perceptions of Artificial Intelligence (AI) Evan A. Schnidman, Prattle}}
This presentation focused on the relative use of AI terminology across central banks and corporations. The main findings were:

- Both central banks and corporations are discussing AI with much greater frequency over the last 3-5 years. This finding was consistent across central banks and corporate asset classes, but some outliers are discussing AI a great deal more than their peers.

- The Bank of England remarks on AI with greater frequency than other central banks. The data also indicates that the Bank of England remains a thought leader among central banks. Over 30 years ago the BOE led global central banks toward increased transparency and more recently BOE personnel have been at the forefront of blockchain research.

- The financial services sector appears to be discussing AI with greater frequency than other sectors, but this finding could be a bit misleading as the number of finance companies are overrepresented in the sample of public US companies. Collectively sectors such as technology, health technology, and electronic technology are discussing AI more frequently than the financial services industry. 

-  Mentions of AI increased more substantially among mid- and small-cap companies than among large-cap companies.

\paragraph{What should researchers do?}
Researchers should further study to determine how actively AI is being deployed and in what ways it is being utilized. A comprehensive AI index indicating which companies are using this technology (as measured by staffing, patents and other measures), as opposed to merely talking about it, could be a valuable path forward to continue this research.


\subsubsection[Coursera Global AI Skill Index Vinod Bakthavachalam, Coursera]{\texorpdfstring{\textit{Coursera Global AI Skill Index} \\ Vinod Bakthavachalam, Coursera}{Coursera Global AI Skill Index Vinod Bakthavachalam, Coursera}}

This presentation focused on the Coursera Global Skills Index (GSI) that draws upon rich data to benchmark 60 countries and 10 industries across Business, Technology, and Data Science skills to reveal skill development trends around the world.

Coursera measures the AI skill proficiency of countries and industries overall and in the related skills of math, machine learning, statistics, statistical programming, and software engineering.

Using a derivative of the Elo algorithm for chess rankings, Coursera measures the skill proficiency of each learner on the platform across these skills based on their performance on relevant graded assessments, adjusting for the difficulty of assessments taken.

- Developed countries tend to place in the top half of the world skill rankings while developing countries rank in the bottom half. The current inequality is likely to increase.

- The authors find that the correlation between AI skills and skills in statistics and machine learning is the highest while the correlation between AI and other skills like software engineering is lower. This is likely because the current weight of AI applications revolves around utilizing machine learning and related statistics concepts today.

\paragraph{What should researchers do?}
Researchers should discover potential new uses of this data, such as tracking country and industry progress in AI skills over time, connecting external data to better understand how learning drives performance, and creating novel hiring signals off of these skill proficiency measures to unlock new career opportunities for people. Related research questions are how these skills can create pathways for future work opportunities.

\subsection{AI for Sustainable Development and Human Rights: Inclusion, Diversity, Human Dignity}
\label{sec:headings4}


\subsubsection[The Problem with Metrics --- what's not captured? Rachel Thomas, University of San Francisco]{\texorpdfstring{\textit{The Problem with Metrics --- what's not captured?} \\ Rachel Thomas, University of San Francisco}{The Problem with Metrics --- what's not captured? Rachel Thomas, University of San Francisco}}
This presentation focused on the inherent problem with metrics. The more a metric is used and the more importance placed on it, the less useful it becomes (due to inevitable efforts to manipulate or game it).  This tension is captured in Goodhart’s Law, \say{When a measure becomes a target, it ceases to be a good measure.}

Any metric is just a proxy for what you really care about.  Examples of this abound:

- The difference between citation counts for a paper vs. actual impact. 

- The number of students taught versus what they understood and internalized.  

- Diversity statistics on the composition of a company versus knowing how employees from underrepresented backgrounds were treated, what opportunities they were given, and if their feedback was listened to. 

- Benchmark tasks versus the suitability of those benchmarks towards real-world applications.

\paragraph{What should researchers do?}
Metrics will always be incomplete, and therefore should be combined with qualitative information. Ensuring that a diverse group of stakeholders is not just included, but that their feedback is incorporated into the project.  Such inclusion remains a more complex endeavour than just checking off a task. An example of a failure in this area is the recent revelation that while Sidewalk Labs (a Google/Alphabet subsidiary) held an Indigenous consultation workshop, 0 of the 14 recommendations that resulted from that consultation were included in the \href{https://www.thestar.com/news/gta/2019/10/25/indigenous-elder-slams-hollow-and-tokenistic-consultation-by-sidewalk-labs.html}{final 1,500-page report} . Despite this, the report mentions the Indigenous consultation numerous times.  One Indigenous elder, Duke Redbird, who participated in the consultation said, \say{It was just shocking that there was kind of a blatant disregard for all of the work that we did. They just wanted to check off a box that says, 'We did Indigenous consultation.'} While metrics can be useful, we must beware of the risks of over-emphasis and potential shortcomings as we work towards healthier outcomes.


\subsubsection[Applying AI for social good Monique Tuin, McKinsey Global Institute]{\texorpdfstring{\textit{Applying AI for social good} \\ Monique Tuin, McKinsey Global Institute}{Applying AI for social good Monique Tuin, McKinsey Global Institute}}
This presentation focused on the analysis of McKinsey library of about 160 use cases where there is evidence that  AI can be applied for social good. For about one-third of use cases in the library, an actual AI deployment in some form was identified.

- AI has the potential to contribute to addressing societal challenges spanning all 17 of the United Nations SDGs. The use cases map to all 17 of the SDGs, supporting some aspects of each one. The highest number of use cases in the library map to SDG 3 (Good health and well-being) and SDG 16 (Peace, justice, and strong institutions).

- To scale AI for social good, bottlenecks and risks will need to be addressed. Scaling up will require overcoming significant bottlenecks, especially to address data accessibility, talent availability, and ‘last mile’ implementation challenges.

\paragraph{What should researchers do?} 
\hfill \break
\hfill \break
- Study AI Risks. Large-scale use of AI for social good entails risks that will need to be mitigated, including improving explainability of AI decisions, \href{https://www.mckinsey.com/featured-insights/artificial-intelligence/tackling-bias-in-artificial-intelligence-and-in-humans?cid=other-eml-alt-mgi-mck&hlkid=ce79db52e43a456c95296dd1acb2500f&hctky=11391579&hdpid=99934294-be26-4aae-9f1d-d3eeb25c1218}{addressing bias}, managing data privacy and security, and considering how AI could be used (or misused) by various actors.

- Grow the list of AI use cases for SDGs. Today, we have limited visibility as to where applications of AI for social good are deployed, in which domains, and by which organizations. Measuring this could help the social impact community direct investment to high potential areas with limited deployment.

- Make better use of structured open data for training and transparent deployment. AI could be used to measure progress towards the SDGs themselves. The UN SDGs consist of 244 indicators measuring global metrics, from the number of individuals living within two kilometers of an all-season road, to damages caused by natural disasters. AI, applied to government data sources, satellite imagery, or user-generated mobile data, could help us better understand the impact of efforts to address the UN SDGs, and support measurement of progress towards the goals, globally and at the national and regional level.


\subsubsection[Ethics and AI: Global News Media Daria Mehra, Quid]{\texorpdfstring{\textit{Ethics and AI: Global News Media} \\ Daria Mehra, Quid}{Ethics and AI: Global News Media Daria Mehra, Quid}}
This presentation focused on analysis related to the volume and evolution of the AI ethics narrative in global news outlets over the past year. The authors’ created a comprehensive boolean search query using global news media data from LexisNexis using the  ethics principles as defined by Harvard University’s \href{https://dash.harvard.edu/bitstream/handle/1/42160420/HLS\%20White\%20Paper\%20Final_v3.pdf?sequence=1&isAllowed=y}{Principled Artificial Intelligence} report. The resulting 3,661 unique articles were then classified into seven topic areas based on language similarity: Framework and Guidelines (32\%), Data Privacy Issues (14\%), Facial Recognition (13\%), Algorithm Bias (11\%), Big Tech Advisory on Tech Ethics (11\%), Ethics in Robotics and Driverless Cars (9\%), and AI Transparency (6.7\%).

\paragraph{What should researchers do?}
These results indicate that while government and international institutions are leading voices in the conversation around setting guidelines and developing ethics frameworks, concerns over data privacy, algorithm bias, and the growing prevalence of facial recognition technologies are hotly debated topics in the public arena. Researchers should develop an open-source monitoring method to track the popularity of AI and ethics narrative to dynamically assess AI opportunities and risks by regions.


\subsubsection[Analyzing Metrics to Track Autonomy in Weapon Systems Marta Kosmyna, Campaign to Stop Killer Robots]{\texorpdfstring{\textit{Analyzing Metrics to Track Autonomy in Weapon Systems} \\ Marta Kosmyna, Campaign to Stop Killer Robots}{Analyzing Metrics to Track Autonomy in Weapon Systems Marta Kosmyna, Campaign to Stop Killer Robots}}

This presentation focused on available evidence on public opinion about autonomous weapons (AWs) and deployed AWs around the world. The Campaign to Stop Killer Robots commissioned three independent polls (Ipsos 2017, Ipsos 2018 , and YouGov 2019) to gauge public opinion on fully autonomous weapons. A \href{https://www.stopkillerrobots.org/2019/11/new-european-poll-shows-73-favour-banning-killer-robots/}{YouGov} poll of ten EU countries, conducted in October 2019, showed 73 percent of European respondents thought their government “should work towards an international ban on lethal autonomous weapons systems.” 

A \href{https://www.ipsos.com/en-us/news-polls/human-rights-watch-six-in-ten-oppose-autonomous-weapons}{2018 global Ipsos poll} surveying 28 countries, showed 61 percent of the public opposed the use of fully autonomous weapons systems. Of those opposed, a majority of respondents cited their reasoning as these weapons would be “unaccountable.” A \href{http://www.openroboethics.org/wp-content/uploads/2015/11/ORi_LAWS2015.pdf}{2015 poll} conducted by the Canadian Open Roboethics Initiative showed 56 percent of global participants thought “lethal autonomous weapons systems should not be developed or used.”

A \href{https://www.sipri.org/publications/2017/other-publications/mapping-development-autonomy-weapon-systems}{2017 report from the Stockholm International Peace Research Institute (SIPRI)} analyzed the development of autonomous weapons systems, how these systems are currently used, and potential options for the regulation and monitoring of their development. These reports rely on open-source information such as news articles, industry and defense guides, scientific publications, company websites, surveys, and press releases.

The presentation made it clear that one of the key challenges in measuring the impact of the classified nature of government information is related to the development of fully autonomous weapons.

\paragraph{What should researchers do?}
To measure how AI relates to lethal autonomous weapons, researchers could track mentions of key-words in digital news sources, such as “AI and warfare,” “AI and policing,” “autonomous weapons,” “killer robots,” “automatic target identification,” etc. Open source research could also track notable statements from public officials on fully autonomous weapons, national defense budgets for AI, as well as the development of national AI policies. The Campaign to Stop Killer Robots keeps a \href{https://www.stopkillerrobots.org/about/}{chronology} of work and tracks events, conferences, and panels related to the topic of fully autonomous weapons. Researchers should also track data related to AWs in developing countries and avoid overemphasis on the development of these weapons by advanced economies.


\subsubsection[Artificial Intelligence in Sub-Saharan Africa Muchiri Nyaggah, Local Development Research Institute (LDRI)]{\texorpdfstring{\textit{Artificial Intelligence in Sub-Saharan Africa} \\ Muchiri Nyaggah, Local Development Research Institute (LDRI)}{Artificial Intelligence in Sub-Saharan Africa Muchiri Nyaggah, Local Development Research Institute (LDRI)}}
This presentation focused on ongoing AI activity in Sub-Saharan Africa (SSA) and ideas to grow the adoption of AI in SSA. The presentation highlighted some of the ways different organizations are investing in AI in SSA. Some examples included:

- Google is investing in AI labs on the continent with the first one already operational in Ghana.

- Microsoft is rolling out a \$100m investment in AI, machine learning, mixed reality development starting in two countries in Africa. 

- NASA, the Australian government, and the Global Partnership for Sustainable Development Data are bringing processed earth observation data to developers working on big data and AI for development in Africa via a scheme called the Africa Regional Data Cube.

\paragraph{What should researchers do?}
Researchers should help release more data and address the following challenges to grow adoption of responsible AI in development across SSA including:

- Capacity: Unless we deal with the low availability of academic programs and their enrollment in Africa, we will have low capacity and be net importers of AI with its accompanying biases. We therefore need to measure the quantity and content of AI and AI-related academic programs in African universities, as well as the growth of faculties and the enrollment/graduation from these programs.

- Enabling environment: A conducive market environment is required to enable innovation, investment, and growth of AI. We therefore need to measure periodically the extent to which critical policy-level enablers are in place and enforced.

- Accountability: If we fail to ensure there’s some level of algorithmic transparency in government we risk perpetuating and strengthening existing inequalities and injustices in the global south. We, therefore, need to track instances of algorithmic injustice and their resolution to hold public sector institutions to account on improvements required, procurement, data use and possibly even restitution.


\subsubsection[Algorithm Appreciation: People prefer algorithmic to human Jennifer Logg, Julia Minson, Don Moore, Harvard Kennedy School and Haas School of Business, UC Berkeley]{\texorpdfstring{\textit{Algorithm Appreciation: People prefer algorithmic to human} \\ Jennifer Logg, Julia Minson, Don Moore, Harvard Kennedy School and Haas School of Business, UC Berkeley}{Algorithm Appreciation: People prefer algorithmic to human Jennifer Logg, Julia Minson, Don Moore, Harvard Kennedy School and Haas School of Business, UC Berkeley}}
This presentation focused on results from eight experiments \cite{logg2019algorithm} suggesting that lay people show “algorithm appreciation.” Specifically, people are more receptive to advice when they think it comes from an algorithm than a person when making estimates and forecasts. People showed “algorithm appreciation” when making estimates about: 
- A visual stimulus (Experiment 1A; N = 202)

- Forecasting the popularity of songs (Experiment 1B; N = 215) and 

- Forecasting romantic matches (Experiment 1C; N = 286)  

Numeracy appears to be a mechanism for reliance on algorithms; those in Experiment 1A who scored higher on a math test \cite{schwartz1997role} relied more on algorithmic advice.This effect was obtained across multiple operationalizations of “human” vs. “algorithmic” advice. 

Interestingly, when the researchers provided judgment and decision making researchers from the society’s mailing list with the survey materials from Experiment 1C (Experiment 1D; N = 119), they predicted the opposite results. Although mTurk participants displayed algorithm appreciation when predicting romantic attraction, researchers expected them to display algorithm aversion.  The idea that people are averse to algorithmic advice is evidently pervasive, even among expert researchers.

Reliance on algorithms was robust in presenting the advisors jointly or separately (Experiment 2; N = 154). However, algorithm appreciation waned when people chose between an algorithm’s estimate and their own (versus an external advisor’s; Experiment 3; N = 403). These tests are important because participants could improve their accuracy by relying on algorithms more than they already do (Experiment 4; N = 671) relative to a normative standard.

Experiment 5 examined how decision makers’ own expertise influenced reliance on algorithms. National security professionals who make forecasts on a regular basis responded differently to algorithmic advice than laypeople did. Experts (N = 70) relied less on algorithmic advice compared to laypeople (N = 301), heavily discounting any advice they received. Although providing advice from algorithms may increase adherence to advice for non-experts, it seems that algorithmic advice often falls on deaf expert ears.

\paragraph{What should researchers do?}
Researchers should explore experimental studies at the intersection of psychology and artificial intelligence. Psychology research already plays an important role in Human-Computer Interaction (HCI) for companies, however, important societal questions on how (and whether) AI shifts human cognitive capabilities, human judgement, and more broadly human perception about societal issues is an open area of research. In this regard, researchers are also exploring anthropomorphic approaches treating AI systems with human interactions.


\subsubsection[Tracking Jobs, Skills and Inclusion in AI through LinkedIn Mar Carpanelli, LinkedIn]{\texorpdfstring{\textit{Tracking Jobs, Skills and Inclusion in AI through LinkedIn} \\ Mar Carpanelli, LinkedIn}{Tracking Jobs, Skills and Inclusion in AI through LinkedIn Mar Carpanelli, LinkedIn}}
This presentation focused on the challenges of measuring how AI and AI skills relate to the future of work. LinkedIn is able to leverage data across roughly 650 million workers, 30 million companies, 23 million jobs, 90 thousand schools, 35 thousand skills, and hundreds of billions of pieces of content. This data can help to build a digital representation of the global economy called the \href{https://economicgraph.linkedin.com/}{Economic Graph}. By mapping every member, company, job, and school, LinkedIn is able to transform the data into insights about the changing world of work that policymakers and researchers can use to complement traditional statistics. Measuring AI Talent is important for the Future of Work because AI is changing the way the global economy operates by (i) transforming jobs, (ii) requiring new skills, and (iii) creating new inclusion challenges.

- \textbf{Transforming jobs:} One of LinkedIn metrics to look at how AI is transforming jobs is the AI hiring rate --- which is an index of how AI talent is being hired across countries over time. Preliminary results of joint research with the Stanford AI Index team reveal that AI hiring is growing across all countries in the sample of over 30 countries, and some of the fastest-growing countries are developed economies, such as Singapore, Australia, Canada, the US, but developing countries are catching fast (incl. India and Brazil).

- \textbf{Requiring new skills:} LinkedIn leveraged member skills to identify the most representative AI skills across occupations in the countries in the sample, their “Skills Genome.” For example, python libraries like Natural Language Toolkit (NLTK) or Machine Learning (ML) are highly prevalent across most countries, but sentiment analysis and text classification are especially prevalent in Singapore and India.  

- \textbf{Inclusion challenges:} One of the many ways to look at inclusion challenges is gender. LinkedIn measured how intensively AI skills are used by different genders across occupations and countries and found that across almost every country the average occupation held by a female is less likely to signal AI skills as compared to the average occupation held by a male. This said, Canada and some European countries (such as Switzerland and France) seem to be ahead in terms of AI skill intensity among females.

\paragraph{What should researchers do?}

These kinds of metrics provide a foundation for a data-grounded discussion about the labor market challenges associated with the emergence of AI and other disruptive technologies, and the policies governments can put in place to both foster AI ecosystems and also tackle those challenges head-on. One example of this in action: by identifying countries with fast growth in AI hiring, researchers can take a closer look into career transitions into (and out of) the industry and identify how AI skills are diffusing across industries. These insights help policymakers update educational programs and also help the private sector invest in relevant training. Similar data and measurements can also shed light on the nature of gender gaps, helping policymakers design interventions aimed at addressing the pipeline gap and the skills gap.


\subsubsection[Global AI Talent Yoan Mantha, ElementAI]{\texorpdfstring{\textit{Global AI Talent} \\ Yoan Mantha, ElementAI}{Global AI Talent Yoan Mantha, ElementAI}}

There is strong evidence that the supply of top-tier AI talent does not meet the demand. Yet there is little visibility on precisely how scarce this talent is or where it is concentrated around the world. This presentation summarized ElementAI’s second survey of the scope and breadth of the worldwide AI talent pool. This research relied on three main data sources. First, to get a picture of the researchers who are moving the field forward, the authors’ reviewed the publications from 21 leading scientific conferences in the field of AI and analyzed the profiles of the authors. Second, the authors analyzed the results of several targeted LinkedIn searches, which showed how many individuals self-report that they have doctorates as well as the requisite skills in different regions around the world. Finally, data was collected from outside reports and other secondary sources to help put the findings in context and better understand the talent pool in a rapidly changing global AI landscape.

The results showed that: 

- 22,400 people published at one or more of the top conferences in the field of machine learning in 2018, up 36\% from 2015 and 19\% from 2017. 

- The number of peer-reviewed publications also rose, up 25\% from 2015 and 16\% from 2017. 

- Women were underrepresented, making up just 18\% of the researchers publishing in these conferences.

- The AI talent pool is highly mobile, with about one-third of researchers (out of the 22,400 published conference authors) working for an employer based in a country other than the one where they received their PhD. 

- About 18\% of the authors who published their work at the 21 conferences included in this survey contributed to research that had a major impact on the overall field as measured by citation counts in the last two years (2017-2018). 

- The countries with the highest number of high-impact researchers (i.e., those within the 18\%) were the United States, China, the United Kingdom, Australia, and Canada.

\paragraph{What should researchers do?}

Researchers should try to fuse multiple data sources together to help them build new proxy indicators of major trends. They should also try to model the geographic aspects of different scientific communities as this can provide information about country-level competitiveness. A more fundamental measurement challenge for researchers is correcting bias from various data sources for accurate nationally representative samples for the AI talent pool. For example, a complementary survey of LinkedIn profiles indicated a total of 36,524 people who qualified as self-reported AI specialists, according to ElementAI’s search criteria. This represents a 66\% increase from the 2018 Global AI Talent report. Nevertheless, the findings in this survey indicate that there has been notable growth and expansion, both in self-reported AI expertise and in the number of authors and scientific papers published at AI conferences, reflecting a field that is dynamic and unmistakably international.

\newpage
\bibliographystyle{plain}
\bibliography{MeasurementAIBib.bib}

\newpage

\chapter{\textbf{\Large Appendix. List of Participants}}


\begin{longtable}{|p{.20\textwidth}|p{.8\textwidth}|}
\hline  
\textbf{Name}   & \textbf{Organization} \\           
\hline                                  \\
Grace Abuhamad  & Element AI            \\
Pedro Avelar    & Federal University of Rio Grande do Sul (UFRGS) \\
Monica Anderson & Pew Research Center   \\
Alessandro Annoni   & European Commission, Joint Research Centre \\
Juan Aparicio   & Siemens               \\
Susan Athey     & Stanford University    \\
Vinod Bakthavachalam    & Coursera       \\
Mohsen Bayati           & Graduate School of Business (GSB), Stanford University    \\
Guy Berger      & LinkedIn, Economic Graph Team \\
El Bachir Boukherouaa   & International Monetary Fund (IMF) \\
Andrei Broder           & Google                \\
Emma Brunskill  & Stanford University   \\
Erik Brynjolfsson       & MIT, and AI Index Steering Committee \\
Tom Campbell    & Future Grasp           \\
Mar Carpanelli  & LinkedIn, Economic Graph Team \\
Danielle Cass   & Workday               \\
Charina Choi    & Global Policy, Emerging Technologies, Google \\
Rita Chung      & Mckinsey Global Institute (MGI)   \\
Jack Clark      & OpenAI, and AI Index Steering Committee   \\
Kyle Clark      & Coursera           \\
Maria de Kleijn & Elsevier          \\
Giuditta De-Prato   & Joint Research Centre, European Commission \\
Devan Desai         & Scheller College of Business, Georgia Institute of Technology  \\
Eileen Donahoe  & Global Digital Policy Incubator, Stanford University’s Cyber Policy Center \\
Tulsee Doshi  & Machine Learning Fairness, Google   \\
Ekkehard Ernst                  & International Labor Organization                                                                                    \\
John Etchemendy                 & HAI Stanford University and AI Index Steering Committee                                                             \\
Agata Foryciarz                 & Stanford University                                                                                                 \\
Alan Fritzler                   & LinkedIn, Economic Graph Team                                                                                             \\
Swetava Ganguli                 & Apple                                                                                                               \\
Ben Gansky                      & Institute for the Future                                                                                            \\
Qiaozi Gao                      & Michigan State University                                                                                           \\
Timnit Gebru                    & Google                                                                                                              \\
Edward Geist                    & RAND                                                                                                                \\
Bernard Ghanem                  & King Abdullah University of Science and Technology                                                                  \\
Ilana Golbin                    & Artificial Intelligence Accelerator, PwC                                                                            \\
Ben Goldhaber                   & ai.metaculus and Parallel Forecast                                                                                  \\
Chris Grainger                  & Amplified.ai                                                                                                        \\
Anil Gupta                      & Smith School of Business, University of Maryland College Park                                                       \\
Gavin Hartnett                  & RAND                                                                                                                \\
Bill Haworth                    & International Finance Corporation (IFC)                                                                             \\
Mario Herger                    & Enterprise Garage Consulting                                                                                        \\
Daniel E. Ho                    & Stanford University                                                                                                 \\
Robert Huschka                  & Association for Advancing Automation and Robotic Industries Assoctiaion                                             \\
Zhao Jin                        & University of Rochester                                                                                             \\
Ramesh Johari                   & Management Science and Engineering, Stanford University                                                             \\
Dan Jurafsky                    & Stanford University                                                                                                 \\
Henry Kautz                     & Information \& Intelligent Systems (IIS), National Science Foundation                                               \\
Matt Kenney                     & Duke University, and AI Index                                                                                       \\
Ashraf Khan                     & International Monetary Fund                                                                                         \\
Marco Konopacki                 & Institute for Technology \& Society of Rio (ITS-Rio)                                                                \\
Marta Kosmyna                   & Campaign to Stop Killer Robots                                                                                      \\
Lars Kotthoff                   & Computer Science, University of Wyoming                                                                             \\
Luis Lamb                       & Department of Theoretical Informatics, Federal University of Rio Grande do Sul (UFRGS)                              \\
Megan Lamberth                  & Technology and National Security Program, Center for a New American Security (CNAS)                                  \\
Burton Lee                      & Stanford University                                                                                                 \\
Xuan Hong Lim                   & Government of Singapore, and World Economic Forum                                                                   \\
Ashley Llorens                  & Intelligent Systems Center, Johns Hopkins University Applied Physics Laboratory (JHU/APL)                           \\
Andrew Lohn                     & RAND Corp.                                                                                                                \\
Jamila Loud                     & Google                                                                                                              \\
Kent Lyons                      & Toyota Research Institute (TRI)                                                                                     \\
Terah Lyons                     & Partnership on AI                                                                                                   \\
Bill MacMillan                  & Prattle                                                                                                             \\
Vikram Mahidhar                 & Genpact                                                                                                             \\
Christos Makridis               & MIT Sloan School of Management                                                                                      \\
Yoan Mantha                     & Element AI                                                                                                          \\
Juan Mateos-Garcia              & Innovation Mapping, NESTA                                                                                           \\
Daria Mehra                     & Quid Inc.                                                                                                           \\
Megan Metzger                   & Global Digital Policy Incubator (GDPi) Program, Stanford University                                                 \\
Saurabh Mishra                  & HAI-AI Index Stanford University                                                                                    \\
Don Moore                       & Haas School of Business, UC Berkeley                                                                                \\
Dewey Murdick                   & Center for Security and Emerging Technology (CSET), Georgetown University                                           \\
Jon Neitzell                    & Anduril Partners                                                                                                    \\
Mark Nelson                     & Stanford University                                                                                                 \\
Muchiri Nyaggah                 & Local Development Institute                                                                                         \\
Lara O'Donnell                  & Element AI                                                                                                          \\
Osonde Osoba                    & RAND                                                                                                                \\
Michael Page                    & Center for Security and Emerging Technology (CSET), Georgetown University                                           \\
Ray Perrault                    & SRI International and AI Index Steering Committee                                                                   \\
Karine Perset                   & Digital Economy and Artificial Intelligence Policy, OECD                                                            \\
Eduardo Plastino                & Accenture                                                                                                           \\
Christopher Potts               & Stanford University                                                                                                 \\
Tamara Prstic                   & Stanford University                                                                                                 \\
Margarita Quihuis               & Peace Innovation Lab, Stanford University                                                                           \\
Anand Rao                       & PwC                                                                                                                 \\
Alex Ratner                     & Stanford University                                                                                                 \\
Chris Re                        & Stanford University                                                                                                 \\
Mark Reidl                      & Georgia Tech 
                                    \\
Maxime Rivest                   & Elsevier                                                                                                            \\
Enrico Robles                   & Endeavor Mexico                                                                                                     \\
Daniel Rock                     & MIT Initiative on the Digital Economy                                                                               \\
Dorsa Sadigh                    & Stanford University                                                                                                 \\
Mehran Sahami                   & Computer Science, Stanford University.                                                                              \\
Emily Sands                     & Coursera                                                                                                            \\
Grigory Sapunov                 & Intento                                                                                                             \\
Peter Sarlin                    & Silo.ai                                                                                                             \\
Konstantin Savenkov             & Intento                                                                                                             \\
Evan Schnidman                  & Prattle                                                                                                             \\
Michael Sellitto                & HAI, Stanford University                                                                                             \\
Luis Serrano                    & Artificial Intelligence, Apple                                                                                      \\
Jake Silberg                    & McKinsey Global Institute                                                                                           \\
Lisa Simon                      & Stanford University                                                                                                 \\
Andrew Smart                    & Google                                                                                                              \\
Ruth Starkman                   & Stanford University                                                                                                 \\
Kostas Stathoulopoulos          & NESTA                                                                                                               \\
Fabro Steibel                   & Institute for Technology \& Society of Rio (ITS-Rio)                                                                \\
Adina Sterling                  & Graduate School of Business, Stanford University                                                                    \\
Robert Stojnic                  & AtlasML                                                                                                             \\
Bledi Taska                     & BurningGlass                                                                                                        \\
Rachel Thomas                   & University of San Francisco (USF) Data Institute                                                                    \\
Ramin Toloui                    & Stanford University                                                                                                 \\
Monique Tuin                    & McKinsey Global Institute (MGI)                                                                                     \\
Christopher Walker              & National Endowment for Democracy                                                                                    \\
Kuansan Wang                    & Microsoft Research (MSR), Outreach Academic Services                                                                \\
Susan Woodward                  & Sandhill Econometrics                                                                                               \\
Chenggang Xu                    & Cheung Kong Graduate School of Business and London School of Economics                                                                            \\
James Zou                       & Stanford University      \\                                               
\hline 
\end{longtable}


\end{document}